\documentclass[11pt,twoside]{article}

\usepackage{graphicx}
\usepackage{asp2010}
\usepackage{bm}

\resetcounters 

\begin{document}

\resetcounters

\title{Third post-Newtonian spin-orbit effect in the gravitational radiation
  flux of compact binaries}
 \author{Luc Blanchet,$^1$ Alessandra Buonanno,$^2$ and Guillaume Faye$^1$
\affil{$^1$Institut d'Astrophysique de Paris,
   UMR 7095 CNRS,\\ Universit\'e Pierre \& Marie Curie, 98$^{\textrm{bis}}$
   boulevard Arago, 75014 Paris, France\\
   $^2$Maryland Center for Fundamental
    Physics \& Joint Space-Science Institute,\\ Department of Physics, 
    University of Maryland, College
    Park, MD 20742, USA}}

\begin{abstract} 
  Gravitational waves contain tail effects that are due to the backscattering
  of linear waves in the curved space-time geometry around the source. The
  knowledge as well as the accuracy of the two-body inspiraling post-Newtonian
  (PN) dynamics and of the gravitational-wave signal has been recently
  improved, notably by computing the spin-orbit (SO) terms induced by tail
  effects in the gravitational-wave energy flux at the 3PN order. Here we sketch
  this derivation, which yields the phasing formula including SO tail effects
  through the same 3PN order. Those results can be employed to improve the
  accuracy of analytical templates aimed at describing the whole process of
  inspiral, merger, and ringdown.
\end{abstract}

\section{Introduction} 
By $2016-2018$, the ground-based gravitational-wave detectors Virgo and LIGO
will be upgraded to such a sensitivity that event rates for coalescing binary
systems will increase by approximately a factor one thousand, making likely
the first detection of gravitational waves from those systems. The search for
gravitational waves from coalescing binary systems and the extraction of
source parameters require a rather accurate knowledge of the waveform of the
incoming signal. The post-Newtonian (PN) expansion is the most powerful
approximation scheme in analytical relativity capable of describing the
two-body dynamics and the gravitational-wave emission of inspiraling compact
binary systems~\citep{Bliving}.

The presence of spin effects adds substantial complexity to the gravitational
waveforms, making it indispensable to include them into search templates. We
have recently improved~\citep{BBF10} the knowledge and accuracy of the
two-body inspiraling dynamics and gravitational-wave signal by computing the
spin-orbit (SO) terms induced by tail contributions due to the
back-scattering of linear waves in the curved space-time geometry around the
source. Here we shall summarize this work, which is the continuation
of~\cite{FBB06} and~\cite{BBF06} where we obtained the 2.5PN
SO contributions in the equations of motion and gravitational-wave energy
flux.

After briefly reviewing the post-Newtonian multipole moment formalism and
discussing relevant properties of tails, we describe how spin effects are
included to our scheme and derive the binary's evolution equations when
black holes carry spins. Next, we investigate the time evolution of the moving
triad and solve the precessing dynamics at the relevant PN order. We then
compute the tails, which depend on the recent past history of the source,
restricting ourselves to quasi-circular adiabatic inspiral. At last, we derive
in the 3PN SO tail effects in the energy flux and in the gravitational
phasing.

\section{Wave generation formalism} 
\label{sec:multipole}

The gravitational waveform $h_{ij}^\mathrm{TT}$, generated by an isolated
source described by a stress-energy tensor $T^{\mu\nu}$ with compact support,
and propagating in the asymptotic regions of the source, is the transverse
trace-free (TT) projection of the metric deviation at the leading-order $1/R$
in the distance to the source. It is parametrized by symmetric trace-free
(STF) mass-type moments $U_L$ and current-type ones $V_L$, referred to as
radiative moments, which constitute observable quantities at infinity from the
source. The general expression of the TT waveform, in a suitable radiative
coordinate system $X^\mu=(c\,T,\mathbf{X})$, reads~\citep{Th80}, when
neglecting terms $\mathcal{O}(1/R^2)$ with $R=\vert\mathbf{X}\vert$
\begin{equation}\label{eq:hijTT}
h^\mathrm{TT}_{ij} = \frac{4G}{c^2R} \,\mathcal{P}^\mathrm{TT}_{ijkl}
\sum^{+\infty}_{\ell=2}\frac{N_{L-2}}{c^\ell\ell !} \biggl[
U_{klL-2} - \frac{2\ell}{c(\ell+1)}\,N_{m}
\,\varepsilon_{mn(k}
\,V_{l)nL-2} \biggr] \, .
\end{equation}
The moments $U_L$ and $V_L$ are functions of the retarded time $T_R\equiv
T-R/c$. The integer $\ell$ refers to the multipolar order and $\mathbf{N}
= \mathbf{X}/R$ is the unit vector pointing from the source to the far
away detector. The tensor $\mathcal{P}^\mathrm{TT}_{ijkl}$ denotes the
transverse-traceless (TT) projector
$\mathcal{P}_{i(k}\mathcal{P}_{l)j}-\mathcal{P}_{ij}\mathcal{P}_{kl}/2$, where
$\mathcal{P}_{ij}=\delta_{ij}-N_iN_j$ is the projector orthogonal to
$\mathbf{N}$. The quantity $\varepsilon_{ijk}$ is the Levi-Civita symbol
such that $\varepsilon_{123}=1$. Round parentheses indicate symmetrization.
After plugging Eq.~(\ref{eq:hijTT}) into the standard expression for the
gravitational-wave energy flux, we get
\begin{equation} \label{eq:flux}
\mathcal{F} = \sum_{\ell = 2}^{+ \infty} \frac{G}{c^{2\ell +1}}
\frac{(\ell+1)(\ell+2)}{(\ell-1) \ell \, \ell! (2\ell+1)!!}\biggl[
 U_L^{(1)} U_L^{(1)} 
+ \frac{4\ell^2}{c^2 (\ell+1)^2} V_L^{(1)} V_L^{(1)}\biggr]\,,
\end{equation}
where the superscript $(1)$ stands for the first time derivative
\citep{Th80}.

In the multipolar-post-Minkowskian formalism, developed by
\cite{BD86,BD88,BD92}, the radiative moments are linked to six sets of
multipole moments characterizing the source, collectively called the
``source'' moments and denoted $I_L$, $J_L$, $W_L$, $X_L$, $Y_L$, $Z_L$. The
relation between $U_L$, $V_L$ and the former quantities encode all the
non-linearities in the wave propagation between the source and the detector.
After being re-expanded in a PN way, they are seen to contain the contribution
of the so-called gravitational-wave tails, due to backscattering of linear
waves onto the space-time curvature associated with the total mass of the
source itself. The expression for $U_L(T_R)$ at the 1.5PN order, where the
tail effects first appear, is
\begin{equation}\label{eq:tails}
U_L(T_R) = I_L^{(\ell)}(T_R) + \frac{2 G M}{c^3} \int_{-\infty}^{T_R}\! \mathrm{d} t \,
 I_L^{(\ell +2)}(t) 
\ln \biggl(\frac{T_R-t}{2\tau'_0} \biggr) 
+\mathcal{O}\Bigl(\frac{1}{c^3}\Bigr)_\mathrm{non-tail}\, .
\end{equation}
Here  $\tau'_0$ is a
freely specifiable time scale~\citep{B95}. A similar equation links
$V_L(T_R)$ to $J_L$. For the present application, $W_L$, $X_L$, $Y_L$ and
$Z_L$, associated to a possible gauge transformation performed at
linear order, play no role. It will be sufficient to consider $I_L$ at 1.5PN
and $J_L$ at 0.5PN order.
The 1.5PN mass moments are given by an integral extending over the mass density
$\sigma \equiv (T^{00}+T^{ii})/c^2$ and the current density $\sigma_i \equiv
T^{0i}/c$ of the matter source~\citep{BBF06}:
\begin{equation} \label{eq:ILS} I_{L} = \int \mathrm{d}^3\mathbf{x}\,
  \biggl[ \hat{x}_L \sigma +
  \frac{1}{2c^2(2\ell+3)}\,\hat{x}_{L}|\mathbf{x}|^2\,\sigma^{(2)} -
  \frac{4(2\ell+1)}{c^2(\ell+1)(2\ell+3)}\,\hat{x}_{iL}
  \,\sigma^{(1)}_{i}\biggr]+\mathcal{O}\Bigl(\frac{1}{c^4}\Bigr)\,.
\end{equation}
The 0.5PN part of $J_L$ is obtained by substituting $\varepsilon_{ab\langle
  i_\ell} x_{L-1\rangle a} \,\sigma_b$ to $\hat{x}_L \, \sigma$ in the
first term.

To control the past behavior of the tail integral in Eq.~(\ref{eq:tails}),
we  assume that, at early times, the source
was formed from a bunch of freely falling particles moving on some
hyperbolic-like orbits, and forming at a later time a gravitationally bound
system by emission of gravitational radiation. This ensures that the integral
$\mathcal{U}_L(T_R) \equiv \int_{-\infty}^{T_R} \mathrm{d} t
\,I^{(\ell+2)}_{L}(t)\,\ln [(T_R-t)/2\tau'_0]$ is convergent. To compute it, it
is convenient to perform the Fourier decomposition
$I_{L}(t) = \int_{-\infty}^{+\infty} \mathrm{d}\Omega/(2\pi)
\,\widetilde{I}_{L}(\Omega)\,e^{-\mathrm{i} \Omega t}$, and commute the tail
integral with the Fourier one. We obtain a closed-form 
expression by resorting to standard mathematical formulae for the integrals of
$e^{-\mathrm{i} \Omega t} \ln t$. The result
reads (with $\textrm{s}(\Omega)\equiv\frac{\Omega}{\vert\Omega\vert}$ and
$\gamma_E$ being the
Euler constant)
\begin{equation}\label{tailsplitres}
\mathcal{U}_L(T_R)  = \mathrm{i}\int_{-\infty}^{+\infty} \frac{\mathrm{d}
  \Omega}{2\pi}\,(-\mathrm{i}\Omega)^{\ell+1}\,\widetilde{I}_{L}(\Omega)\,
e^{-\mathrm{i}\,\Omega\,T_R} \left[\frac{\pi}{2}\textrm{s}(\Omega) +
  \mathrm{i}\Bigl(\ln(2\vert\Omega\vert\tau_0)+\gamma_\mathrm{E}\Bigr)\right]\,.
\end{equation}

\section{Applications to spinning binaries}
\label{sec:appl}

\subsection{Spin vectors for point-like objects} \label{sec:appl1}

Following our previous works~\citep{FBB06, BBF06}, we base our
calculations on the model of point-particles with spins~\citep{Mathisson37,
  Papa51spin, Tulc2, BI80}. The stress-energy tensor $T^{\mu\nu}$ of a
system of spinning particles is the sum of a monopolar piece, made of Dirac
delta-functions, plus the dipolar or spin piece, made of gradients of
delta-functions:
\begin{equation}\label{eq:Tmunu}
T^{\mu\nu} =
c^2 \sum_A
\int_{-\infty}^{+\infty} \! \mathrm{d}\tau_A \biggl\{ p_A^{\mu} u_A^{\nu} 
\frac{\delta^{(4)}
(x-y_A)}{\sqrt{-g_A}} 
-\frac{1}{c}\,\nabla_\rho\biggl[ S_A^{\rho\mu}\,u_A^{\nu} \frac{\delta^{(4)}
(x-y_A)}{\sqrt{-g_A}} \biggr] \biggr\} \,,
\end{equation}
where $\delta^{(4)}$ is the four-dimensional Dirac function, $x^\mu$ is the
field point, $y_A^\mu$ is the world-line of particle $A$, $u_A^\mu=\mathrm{d}
y_A^\mu/(c \mathrm{d}\tau_A)$ is the four-velocity such that
$g^A_{\mu\nu}u_A^\mu u_A^\nu=-1$ (with $g^A_{\mu\nu}\equiv g_{\mu\nu}(y_A)$
denoting the metric at the particle's location), $p_A^\mu$ is the linear
momentum of the particle, and $S_A^{\mu\nu}$ denotes its antisymmetric spin
angular momentum. Our spin variables has been rescaled by a factor $c$
($S^{\mu\nu} = c \, S^{\mu\nu}_\mathrm{true}$) so as to have a non-zero
Newtonian limit for compact objects. Imposing that the antisymmetric part of
$T^{\mu\nu}$ should vanish yields the covariant equations of evolution for the
spin, while the covariant equations of motion follow from the conservation
relation $\nabla_\nu T^{\mu\nu} = 0$ \citep{Papa51spin}.

In order to fix unphysical degrees of freedom associated with an arbitrariness
in the definition of $S^{\mu\nu}$ in the case of point particles (due to the
freedom in the choice of the location of the center-of-mass worldline within
the extended bodies), we adopt the covariant supplementary spin condition
$S_A^{\mu\nu}\,p^A_\nu = 0$~\citep{Tulc2}. We restrict the computation to
linear terms in the spins, neglecting $\mathcal{O}(S_A^2)$. Then the 
mass $m_A \equiv (-p_A^\mu p^A_\mu)^{1/2}$ is conserved and
 $S^{\mu\nu}_A$ is parallel transported along the worldline of body $A$. As is
standard, we introduce a spin vector variable $S_A^i$ with constant magnitude
\citep[see e.g.,][]{K95}, for instance by projecting $S_A^{\mu\nu}$ onto some
orthonormal basis of the $A$-particle rest space and taking the Hodge dual
with $\varepsilon_{ijk}$.

The post-Newtonian expressions of the densities $\sigma$, $\sigma_i$ entering
the source moments~(\ref{eq:ILS}), as well as the stress density
$\sigma_{ij}\equiv T^{ij}$, are obtained iteratively from the truncated
components of the stress-energy tensor~(\ref{eq:Tmunu}). They are used in turn
to compute the metric at the next iteration step of the perturbative scheme.

Inserting the former matter densities into Eq.~(\ref{eq:ILS}) (or its
counterpart for $V_L$), we see that spins arise at 1.5PN order in $I_L$ and
0.5PN order in $J_L$. By virtue of Eq.~(\ref{eq:flux}), leading SO terms in the
gravitational-wave flux are thus seen to be of half-integer PN order. However,
due to the $1/c^3$ factor in front of the tail integrals, as shown  in
Eq.~(\ref{eq:tails}), terms of integer orders may also appear:
(i) when the source moments are contained in the tail integrands, or (ii) when
they are multiplied by non-spin tail integrals. The leading 3PN terms coming
from $U_{ij}$ and $V_{ij}$ are precisely those we want to compute.

\subsection{1.5PN dynamics with spin-orbit effects} \label{sec:appl2}

To reduce the accelerations generated by the time differentiations of $I_{ij}$
and $J_{ij}$ and, most importantly, to find the time dependence of
$I_{ij}^{(4)}$ and $J_{ij}^{(4)}$ required for calculating integrals such as
the one in Eq~(\ref{eq:tails}), we shall rely on the 1.5PN dynamics. The
corresponding PN equations of evolution at this order are derived by inserting
the 1.5PN metric into the covariant equations and discarding
$\mathcal{O}(1/c^4)$ remainders. Going to the center-of-mass frame, we denote
by $\mathbf{x}=\mathbf{y}_1-\mathbf{y}_2$ the relative position of the
particles (and $r=\vert \mathbf{x}\vert$,
$\mathbf{v}=\mathrm{d}\mathbf{x}/\mathrm{d} t$). Following~\cite{K95} we
introduce an orthonormal moving triad $\{\mathbf{n},\bm{\lambda},
\bm{\ell}\}$ defined by
$\mathbf{n}=\mathbf{x}/r$,
$\bm{\ell}=\mathbf{L}_\mathrm{\scriptscriptstyle
  N}/\vert\mathbf{L}_\mathrm{\scriptscriptstyle N}\vert$, where
$\mathbf{L}_\mathrm{\scriptscriptstyle N}\equiv m \nu
\,\mathbf{x}\times\mathbf{v}$ denotes the Newtonian orbital angular momentum
(with $m=m_1+m_2$ and $\nu = m_1 m_2/m^2$), and $\bm{\lambda}=\bm{\ell} 
\times\mathbf{n}$. For general orbits,
the time derivatives of those vectors may be written in the form
\begin{equation} \label{eq:triad_evolution}
  \frac{\mathrm{d}\mathbf{n}}{\mathrm{d} t} = \omega
  \, \bm{\lambda} \, , \qquad 
  \frac{\mathrm{d} \bm{\lambda}}{\mathrm{d} t} = - \omega
  \,\mathbf{n} - \omega_\mathrm{prec} \,\bm{\ell} \, , \qquad  
  \frac{\mathrm{d} \bm{\ell}}{\mathrm{d} t} = 
\omega_\mathrm{prec} \,  \bm{\lambda} \, .
\end{equation}
The angular frequency $\omega$ at which the separation $\mathbf{n}$ rotates in
the instantaneous orbital plane is the so-called orbital frequency, while the
third equation~(\ref{eq:triad_evolution}) defines the precessional frequency
$\omega_\mathrm{prec}$, which gives thus the variation of $\bm{\ell}$ in the
direction of $\bm{\lambda}$.

Instead of writing the equations of motion in the form of a link between
$d\mathbf{v}/dt$ and the kinematic variables, we give the expressions
of $\omega$, $\omega_\mathrm{prec}$. Focusing from now on to the case of
quasi-circular motion, for which $r$ is constant apart from the 2.5PN secular
radiation damping, we have~\citep[see e.g.,][]{FBB06}
\begin{eqnarray}
& &\omega^2 = \frac{G\,m}{r^3}\bigg\{1 + \gamma\left(-3+\nu\right) +
\gamma^{3/2}\left(-5 s_\ell-3 \delta \sigma_\ell\right)\biggr\}
+\mathcal{O}\Bigl(\frac{1}{c^4}\Bigr)\,, \label{eq:omega} \\
& &\omega_\mathrm{prec} = - \omega\,\gamma^{3/2}\Bigl(7 s_n + 3
\delta\sigma_n\Bigr)+ \mathcal{O}\Bigl(\frac{1}{c^4}\Bigr)\,,
\end{eqnarray}
where the dimensionless spin variables are defined by $\mathbf{s} =
(\mathbf{S}_1 + \mathbf{S}_2)/(G m^2)$ and $\mathbf{\sigma} =
(\mathbf{S}_2/m_2 - \mathbf{S}_1/m_1)/(G m)$, whereas $\delta \equiv
X_1 - X_2 $, $s_\ell\equiv \mathbf{s}\cdot\bm{\ell}$ and
$s_n\equiv \mathbf{s}\cdot\mathbf{n}$; $\gamma\equiv G m/(r c^2)$ is
the PN parameter. The spin vectors $\mathbf{S}_A$ satisfy the
usual-looking precession equations
\begin{equation}\label{eq:precession_equations}
  \frac{\mathrm{d} \mathbf{S}_A}{\mathrm{d} t} = \mathbf{\Omega}_A
  \times \mathbf{S}_A\,, \qquad \textrm{with}~~ \mathbf{\Omega}_1 =
  \omega\,\gamma \,\biggl[ \frac{3}{4} + 
  \frac{\nu}{2}  -\frac{3}{4} \delta
  \biggr]\,\bm{\ell} + \mathcal{O}
  \Bigl(\frac{1}{c^4}\Bigr) \,, 
\end{equation}
showing that the spins, at the 1.5PN order, precess around the direction of
$\bm{\ell}$, and at the quasi-constant rate
$\Omega_A=\vert\mathbf{\Omega}_A\vert$. To obtain $\mathbf{\Omega}_2$
we simply have to change $\delta$ into $-\delta$.

\section{Computation of the tails}
\label{sec:comput}

The only time dependence in the source moments composing the tail integrals
occurs through the triad vectors and the spins at the 1.5PN order. We adopt
the following strategy to integrate Eqs.~(\ref{eq:triad_evolution}): (i) we
express the three moving triad vectors in terms of the Euler angles that link
the latter triad to some orthonormal fixed triad whose third vectors points to
the direction of the total angular momentum at some time $t_0$; (ii) we
express the appropriate brick combinations of those angles as a function of
$S_n^A$, $S_\lambda^A$, $S_\ell^A$ and $\phi \equiv \int \mathrm{d}t\,
\omega(t)$; (iii) we solve for the precession
equations~(\ref{eq:precession_equations}) in the triad basis and insert the
results into the formulae we got for $\mathbf{n}$, $\bm{\lambda}$, $\bm{\ell}$
in the previous step.

Although the binary continuous spectrum of frequencies of $I_L(t)$ should
contain all orbital frequencies at any epoch in the past, say $\omega(t)$ with
$t\leq T_R$, differing from the current orbital frequency $\omega(T_R)$ due to
gravitational radiation damping, we can actually compute the tail integral by
considering only the \textit{current} frequency $\omega(T_R)$, modulo small
error terms of negligible order $\mathcal{O}(\ln c/c^5)$. Similarly, we can
regard the precession frequencies $\mathbf{\Omega}_A(t)$ of the two spins as
constant in the calculation. Thus, we may replace $\phi(t)$ by $\omega(t-t_0)
+ \phi(t_0)$ and $\Omega_A(t)$ by $\Omega_A(T_R)$ in the tail integrand. This
takes the form of a sum of complex exponentials of the type $e^{\mathrm{i} \,
  \omega_{npq} t}$, with $\omega_{npq}=n\,\omega+p\,\Omega_1+q\,\Omega_2$,
times certain coefficients $A_{L}^{npq}$. It is then straightforward to take
the Fourier transform of the source moments, e.g. $I_L(t)$, and apply
formula~(\ref{tailsplitres}), which leads to
\begin{equation}\label{taildecomp}
\mathcal{U}_{L}(T_R) =  \sum_{n,p,q}  \mathrm{i}\,
  A_{L}^{npq}(-\mathrm{i}\,\omega_{npq})^{\ell+1}
e^{-\mathrm{i} \,\omega_{n pq}T_R}
\left[\frac{\pi}{2} \textrm{s}(\omega_{n pq}) + \mathrm{i}
    \Bigl(\ln(2\vert\omega_{n pq}\vert\tau_0)+
\gamma_\mathrm{E}\Bigr)\right]\,.
\end{equation}
\section{Energy flux and orbital phasing} 

Inserting the explicit expressions for $U_{ij}$ and $V_{ij}$ into the energy
flux~(\ref{eq:flux}) yields the net SO tail contribution at 3PN order, in
terms of the gauge-invariant PN parameter $x\equiv(G m \omega/c^3)^{2/3}$ and
the spins. The SO terms have to be included in either the instantaneous moment
in front of the tail integrals, or in the tail integrals themselves. This
gives several ``direct'' SO contributions coming from tails at relative
1.5PN order (for the mass quadrupole tail) or 0.5PN order (for the current
quadrupole tail) which are then added together. In addition there is the
crucial contribution due to the reduction to circular orbits of the standard
(non-spin) tail integral at 1.5PN order, for which the relation between the
orbital separation $r$ and the orbital frequency $\omega$ [as given by the
inverse of Eq.~(\ref{eq:omega})] provides a supplementary SO term at relative
1.5PN order, which contributes \textit{in fine} at the same 3PN level as the
``direct'' SO tail terms. Finally we obtain 
\begin{equation}\label{3PNspintail}
\delta\mathcal{F} =\frac{32}{5} \frac{c^5}{G}\,x^8\,\nu^2 \Bigl[ - 16 \pi
  \,s_\ell - \frac{31\pi}{6}\,\delta\,\sigma_\ell\Bigr]\,,
\end{equation}

Using an energy balance argument, we finally equate the averaged evolution of
the binding energy of the binary reduced for quasi-circular orbits
$\mathrm{d}E(x,s_\ell,\sigma_\ell)/\mathrm{d}t$~\citep[see e.g.,][]{BBF06} to
$-\mathcal{F}(x,s_\ell,\sigma_\ell)$, where $\mathcal{F}$ is the sum
of the 2PN SO flux obtained by~\cite{K95}, the 3PN non-spin contribution
investigated by~\cite{BFIJ02, BDEI04} and the correction~(\ref{3PNspintail})
we just calculated. The time derivative applied to $E$ acts only to the
frequency variable since, as we have checked, $s_\ell$ and $\sigma_\ell$ are
secularly conserved (neglecting SS contributions). We find the following
secular variation of frequency, denoted $\dot\omega$ for simplicity, at the
3PN order including SO effects:
\begin{eqnarray}
&& \frac{\dot{\omega}}{\omega^2} = \frac{96}{5}\,\nu\,x^{5/2} \bigg\{ 1 +x
\Bigl(-\frac{743}{336}-\frac{11}{4}\nu \Bigr) + x^{3/2} \Bigl(
4\pi-\frac{47}{3}s_\ell -\frac{25}{4} \delta \sigma_\ell
\Bigr) \nonumber\\ &&\qquad+x^2 \Bigl( \frac{34103}{18144}+\frac{13661}{2016}\nu
+\frac{59}{18}\nu^2 \Bigr) \nonumber\\ && \qquad + x^{5/2} \Bigl(
-\frac{4159}{672} \pi - \frac{5861}{144} s_\ell - \frac{809}{84}
\delta \sigma_\ell + \nu \Bigl[ - \frac{189}{8}\pi +
  \frac{1001}{12} s_\ell + \frac{281}{8} \delta\sigma_\ell
  \Bigr] \Bigr) \nonumber\\ &&\qquad + x^{3} \Bigl(\frac{16447322263}{139708800}+
  \frac{16}{3}\pi^2-\frac{1712}{105}\gamma_\mathrm{E} - \frac{856}{105} \ln
  (16\,x) - \frac{188\pi}{3} s_\ell-\frac{151\pi}{6}
  \delta \sigma_\ell \nonumber\\ && \qquad \qquad + \nu
  \Bigl[-\frac{56198689}{217728} + \frac{451}{48}\pi^2 \Bigr] 
+ \frac{541}{896}\nu^2 - \frac{5605}{2592}\nu^3
\Bigr) \bigg\}\,.
\end{eqnarray}
Integrating this, we get the  ``carrier'' phase $\phi(t) = \int\! \omega
\, dt$. The total phase $\Phi= \phi - \alpha + \mathcal{O}(1/c^4)$, where
$\alpha$ is the precessional phase, can be computed numerically or
manually.

\bibliography{LISAsymp_2012_Faye}
\bibliographystyle{asp2010}

 \end{document}